\documentclass[aps,prd,10pt,twocolumn,superscriptaddress,floatfix,nofootinbib,amsmath,amssymb,altaffilletter,preprintnumbers]{revtex4-1}
\usepackage[colorlinks=true,pdfstartview=FitV,breaklinks=true]{hyperref}

\usepackage{bm}
\usepackage{graphicx}
\usepackage{multirow}
\usepackage{soul}
\usepackage{amsmath}
\usepackage{fontawesome}
\usepackage{mathrsfs}
\usepackage{amssymb}
\usepackage{yfonts}
\usepackage{cancel}
\usepackage{color}
\usepackage{xspace}
\usepackage{url}
\usepackage{verbatim}
\usepackage{mathtools}
\usepackage{upgreek}
\usepackage{units}
\usepackage{siunitx}
\usepackage{amstext}
\usepackage[sc]{mathpazo}
\usepackage{booktabs}
\usepackage{tabulary}
\usepackage{tabularx}
\usepackage{etoolbox}
\usepackage[version=4]{mhchem}
\usepackage[utf8]{inputenc}
\usepackage[dvipsnames,table]{xcolor}
\newcolumntype{Y}{>{\centering\arraybackslash}X}
\hypersetup{urlcolor=BlueViolet,
	    citecolor=Plum,
	    linkcolor=PineGreen}
\usepackage[official]{eurosym}
\definecolor{lightgray}{rgb}{0.9,0.9,0.9}	    
\definecolor{green}{rgb}{0,0.5,0}
\definecolor{red}{rgb}{1,0,0}
\definecolor{blue}{rgb}{0,0,0.5}

\newcommand{\dbd}[2]{\ifmmode \frac{\textrm{d}#1}{\textrm{d}#2}\else $\textrm{d}#1/\textrm{d}#2$\fi}
\newcommand{\pbp}[2]{\ifmmode \frac{\partial#1}{\partial#2}\else $\partial#1/\partial#2$\fi}

\DeclareMathAlphabet{\mathpzc}{OT1}{pzc}{m}{it}
 \newcommand{\eV}{\text{e\kern-0.15ex V}\xspace}

 \newcommand{\TeV}{\text{T\kern-0.1ex \eV}\xspace}

\DeclareMathAlphabet{\mathpzc}{OT1}{pzc}{m}{it}

\newcommand{\be}{\begin{equation}}
\newcommand{\ee}{\end{equation}}
\newcommand{\bea}{\begin{eqnarray}}
\newcommand{\eea}{\end{eqnarray}}

\begin{document}

\title{Cornering the axion with $CP$-violating interactions}

\author{Ciaran A. J. O'Hare}\email{ciaran.ohare@sydney.edu.au}
\affiliation{ARC Centre of Excellence for Dark Matter Particle Physics,
School of Physics, The University of Sydney, Physics Road, NSW 2006 Camperdown, Sydney, Australia}%
 
\author{Edoardo Vitagliano}\email{edoardo@physics.ucla.edu}\affiliation{ Department  of  Physics  and  Astronomy,  University  of  California,  Los  Angeles,  California,  90095-1547,  USA}

\date{\today}
\smallskip
\begin{abstract}
Besides $CP$-preserving interactions, axions and axion-like particles may also have small $CP$-violating scalar Yukawa interactions with nucleons and electrons. Any such interaction will generate macroscopic monopole-dipole forces which can be searched for experimentally. When the best experimental limits on scalar interactions are combined with stellar energy-loss arguments constraining pseudoscalar interactions, strong bounds can be set on $CP$-violating axion couplings which almost intersect the expectation for QCD models. Over the years, both astrophysical and laboratory tests have improved. We provide a much-needed up-to-date compilation of these constraints, showing improvements in some regions of parameter space by factors between 40 and 130. We advocate experimental opportunities, without astrophysical or dark-matter assumptions, to track down the axion in the lesser-explored corners of its parameter space.\,\smallskip\smallskip \href{https://cajohare.github.io/AxionLimits/}{\large\faGithub}

\end{abstract}
\maketitle
\section{Introduction}\label{sec:intro}
The axion is a beyond-the-Standard-Model (SM) pseudoscalar, originally appearing as a consequence of Peccei and Quinn's solution to the strong $CP$ problem of quantum chromodynamics (QCD)~\cite{Peccei:1977hh, Peccei:1977ur, Weinberg:1977ma, Wilczek:1977pj, Kim:2008hd}. As the pseudo-Nambu-Goldstone boson of a new spontaneously broken U(1), the so-called ``QCD axion'' can be engineered with extremely weak couplings to the SM if the symmetry-breaking scale $f_a$ is large. The effective field theory for the axion can be expressed solely in terms of $f_a$, which is inversely proportional to a small mass $m_a$, generated by mixing with the SM mesons. Nevertheless, several UV completions have been devised, such as the popular KSVZ~\cite{Kim:1979if,Shifman:1979if}  and DFSZ~\cite{Dine:1981rt,Zhitnitsky:1980tq} models (see Ref.~\cite{DiLuzio:2020wdo} for a recent review).

In the last decade, efforts to search for the axion have rapidly accelerated. Axions have been shown to be a very viable candidate for the dark matter which dominates the mass budget of the Universe~\cite{Abbott:1982af,Dine:1982ah,Marsh:2015xka}; a motivation that has driven at least part of the axion's recent surge in popularity. Certainly, the aesthetic draw of a particle which solves two problems simultaneously makes it an attractive candidate to test. In the absence of any accidental cancellations, the axion should possess small derivative couplings to fermions, facilitating a large number of tests in both laboratory experiments and astrophysical environments. Some of these tests can rely on the axion comprising galactic dark matter~\cite{Hagmann:1990tj,DePanfilis:1987dk,Asztalos:2010,Brubaker:2016ktl,Ouellet:2018beu,TheMADMAXWorkingGroup:2016hpc,Lawson:2019brd,Alesini:2019ajt,Alesini:2017ifp,Goryachev:2017wpw,McAllister:2018ndu,Du:2018uak,Braine:2019fqb,Boutan:2018uoc,Lee:2020cfj,Zhong:2018rsr,Gramolin:2020ict,Foster:2020pgt,Darling:2020uyo,Darling:2020plz,Thorpe-Morgan:2020rwc,Regis:2020fhw}, or they can be simply a test for the axion's existence as a new particle~\cite{Ballou:2015cka, DellaValle:2015xxa, Bahre:2013ywa,Ehret:2010mh,Betz:2013dza,Semertzidis:1990qc,Moriyama:1998kd,Inoue:2008zp,Anastassopoulos:2017ftl}. A recent review of experimental probes of axions described these in more detail~\cite{Irastorza:2018dyq} 

Although the QCD axion can be defined by one parameter, there will always be $\mathcal{O}(1)$ differences in the axion's various coupling constants depending on the specific model. Therefore it is sensible to set experimental bounds in the broader context of axion-like particles (ALPs), in which the proportional relationships between the axion mass and its couplings are not enforced. The dimensionless $\mathcal{O}(1)$ coupling constants of QCD axion models (i.e.~the ones that solve the strong $CP$ problem) delineate a band in these plots. However models outside of this QCD band are increasingly considered to be interesting in their own right: most notably in the context of some string theories, which are said to populate an ``axiverse''~\cite{Masso:1995tw, Masso:2002ip, Ringwald:2012hr, Ringwald:2012cu, Arvanitaki:2009fg, Cicoli:2012sz, Jaeckel:2010ni} of light to ultralight ALPs. In this article, we adopt the increasingly common (though somewhat unhelpful) usage of the term ``axion'' to refer to any new light pseudoscalar that couples with the same interactions as the true QCD axion.

Experimentally speaking, one of the appealing properties of the axion is that it can mediate macroscopic dipole-dipole forces. These are spin-dependent forces between bodies with some net polarization. Dipole-dipole forces are generated via the axion's generic pseudoscalar couplings and have inspired several experimental campaigns recently~\cite{JacksonKimball:2017elr, Garcon:2019inh, Wu:2019exd}. But as well as these $CP$-even interactions, we have reason to believe that there could be $CP$-violating scalar interactions between the axion and fermions as well~\cite{Moody:1984ba, Georgi:1986kr, Pospelov:1997uv, Pospelov:2005pr}, even though the axion was introduced as part of a solution to explain the \emph{absence} of $CP$ violation in QCD. These need not originate from beyond-the-SM; for example any $CP$ violation coming from the weak sector through the CKM matrix would shift the axion's vacuum expectation value (VEV) and create $CP$-violating Yukawa interactions between the axion and nucleons.

The possibility of scalar axion-nucleon interactions is an intriguing prospect from an experimental standpoint. They would mediate both monopole-monopole (spin-independent), as well as monopole-dipole forces (between spin-polarized and unpolarized bodies, sometimes called spin-mass forces). The discovery of any such forces would be groundbreaking, so are highly sought after. A monopole-monopole interaction, for instance, would lead to scale-dependent departures from firmly established gravitational physics like Newton's inverse-square law and the weak equivalence principle (WEP); see Ref.~\cite{Tino:2020nla} for a recent review. Tests of these laws are important in the exploration of possible modifications of gravity in general~\cite{Perivolaropoulos:2019vkb}. Hence constraints have improved considerably in recent decades with the use of Casimir measurements~\cite{Sushkov:2011md,Chen:2014oda}, microcantilevers~\cite{Geraci:2008hb}, torsion-balance experiments~\cite{Kapner:2006si,Lee:2020zjt,Smith:1999cr,Smith:1999cr,Yang:2012zzb,Tan:2020vpf,Hoskins:1985tn}, and satellite-borne accelerometers~\cite{Berge:2017ovy}. 
Monopole-dipole interactions, on the other hand, can also be searched for with torsion-balance techniques, if one of the masses is spin-polarized ~\cite{Crescini:2016lwj, Crescini:2017uxs, Wineland:1991zz, Lee:2018vaq, Hoedl:2011zz, Terrano:2015sna}; or by searching for the spin depolarization of nucleons when exposed to surrounding bulk matter~\cite{Petukhov:2010dn}. See Ref.~\cite{Safronova:2017xyt} for a review of new physics searches with atoms and molecules.

Laboratory experiments like the ones mentioned above typically test for forces characterized by a range $\lambda$. To reinterpret results of these experiments in the context of axions we use the fact that the range of the axion-induced force is given by the inverse of its mass, $\lambda=1/m_a$.\footnote{We use natural units with $\hbar=c=1$.} Laboratory experiments are competitive down to the $\sim$0.1~$\upmu$m scale, or masses below an eV or so. For higher masses, the experimental limits are superseded by bounds obtained invoking stellar cooling arguments~\cite{Grifols:1986fc, Grifols:1986fc, Raffelt:1994ry, Hardy:2016kme, Carenza:2019pxu,Carenza_2020, Beznogov:2018fda}. As was pointed out by Raffelt in 2012~\cite{Raffelt:2012sp}, the combination of the best experimental bounds on scalar interactions can be multiplied by the best astrophysical bounds on pseudoscalars, resulting in a limit on scalar-pseudoscalar interactions that is better than all other searches devoted to this coupling. It is challenging for the purely experimental monopole-dipole searches to be competitive with this combination. Particles with pseudoscalar couplings can be produced relatively easily in stellar environments, but spin interactions in the lab are in competition with other magnetic interactions, making them difficult to observe. Therefore, despite the abundance of published limits, no experiment has successfully broken through into the band of couplings expected for QCD models, though a few have been proposed. The planned experiment ARIADNE~\cite{Arvanitaki:2014dfa}, has been projected to reach the QCD band for nucleon-nucleon monopole-dipole interactions. For electron-nucleon interactions, a similar experiment QUAX-$g_p g_s$~\cite{Crescini:2016lwj} has been proposed and has already published a limit~\cite{Crescini:2017uxs}. However the experiment will need to extend the sensitivity of its resonant mode considerably to reach the allowed QCD models.

The goal of this article is to find the most competitive and up-to-date laboratory and astrophysical bounds on monopole-monopole and monopole-dipole forces, and use them to compile the most stringent limit on the axion's $CP$-violating couplings. We begin in Sec.~\ref{sec:axions} by reviewing some of the mathematical details of the axion's various $CP$-violating and $CP$-conserving couplings to fermions. Then in Sec.~\ref{sec:astrolimits} we present an up-to-date summary of astrophysical bounds on those couplings, and the axion mass. In Sec.~\ref{sec:baryon}, we compile the most competitive experimental limits on the scalar nucleon interaction. Then, in Sec.~\ref{sec:spinbulk} we present constraints on the combination of the scalar$\times$pseudoscalar couplings for axion-mediated forces between electrons and nucleons, and compare them with experimental monopole-dipole searches. Finally, we conclude with some cautionary remarks about combining astrophysical and laboratory bounds in Sec.~\ref{sec:astrolabbounds}, before summarizing in Sec.~\ref{sec:conclusions}.

\section{Axion couplings}\label{sec:axions}
The characteristic property of the axion is its relation $m_a f_a\sim m_\pi f_\pi$ between the axion's mass $m_a$ and decay constant $f_a$, and those of the pion: $m_\pi$ and $f_\pi$. The most recent lattice QCD calculations give the numerical relationship~\cite{Borsanyi:2016ksw,Gorghetto:2018ocs},
\begin{equation}
 m_a= 5.7 \times 10^{-3}\,\mbox{eV} \,\frac{10^9~\mbox{GeV}}{f_a} \, .
\end{equation}
The axion has a wide range of possible couplings. Here we only explore the standard $CP$-conserving and $CP$-violating interactions between the axion, $a$, and fermions $\psi$,
\begin{equation}\label{eq:lagrangian}
\mathcal{L} \supset -a \sum_{\psi} g_{p}^\psi \left(i \bar{\psi} \gamma^{5} \psi\right) -a \sum_{\psi} g^\psi_s(\bar{\psi} \psi) \, .
\end{equation}
The first sum involves the $CP$-conserving terms which have been rewritten from their derivative coupling form $\partial_\mu a \, \bar\psi_e\gamma^\mu\gamma_5\psi_e$ into a pseudoscalar form. 

The $CP$-conserving and $CP$-violating vertices can be combined in three different ways~\cite{Moody:1984ba}, producing, respectively, a monopole-monopole potential
\begin{equation} \label{eq:monomono}
V=-\frac{g_{s_1}g_{s_2}}{4\pi r}\, e^{-r/\lambda}\, ,
\end{equation}
(where $\lambda=1/m_a$ is the Compton wavelength of the axion) a monopole-dipole potential
\begin{equation}
 V_{sp}=\frac{ g_s g_p}{8 \pi m_\psi}\left( \frac{1}{r \lambda}+\frac{1}{r^2}\right) e^{-\frac{r}{\lambda}} \left(\hat \sigma \cdot \hat r \right) \, ,
\end{equation}
and a \textit{spin-dependent} dipole-dipole potential
 \begin{align}
 V_{pp}=\frac{1}{16 \pi}\frac{ g_{p_1} g_{p_2}}{m_{\psi_1} m_{\psi_2}} \left[\left(\hat \sigma_1 \cdot \hat \sigma_2\right)\left( \frac{1}{r^2 \lambda}+\frac{1}{r^3}\right) e^{-\frac{r}{\lambda}}\right.\\ \nonumber
- \left.\left(\hat \sigma_1 \cdot \hat r \right)\left(\hat \sigma_2 \cdot \hat r\right) \left( \frac{1}{r \lambda^2}+\frac{3}{r^2 \lambda}+\frac{3}{r^3}\right)  e^{-\frac{r}{\lambda}}\right] \, ,
 \end{align}
 where $\hat{\sigma}_{1,2}$ are unit vectors in the directions of the fermion spins.
It is worth noticing that the derivative coupling and the pseudoscalar one are not completely equivalent, as shown, for example, in the discussion contained in Chapter 14 of Ref.~\cite{Raffelt:1996wa}. Any Feynman diagram including more than one derivative vertex is not equivalent to a Feynman diagram with pseudoscalar vertices substituting the derivative ones. Therefore, the potential due to double axion exchange is different in the two cases: a pseudoscalar coupling can produce a $\propto r^{-3}$ \textit{spin-independent} potential between unpolarized bodies, while in the derivative coupling case the spin-independent term is $\propto r^{-5}$~\cite{Ferrer:1998ue}. Therefore, the constraints on pseudoscalar couplings to nucleons are stronger than the ones on derivative couplings; see e.g.~Refs.~\cite{Adelberger:2003ph,Mostepanenko:2020lqe}.
 
We assume the following relationship for the pseudoscalar couplings, which defines a band after choosing a suitable range for the dimensionless $\mathcal{O}(1)$ coupling constants $C_{a\psi}$:
\begin{equation}
 g^\psi_p= \frac{C_{a\psi} m_\psi}{f_a}= 1.75\times 10^{-13} \, C_{a\psi} \,   \left( \frac{m_\psi}{1~{\rm GeV}}\right)\left( \frac{m_a}{1\, \upmu {\rm eV}}\right) \, ,
\end{equation}
where $m_\psi$ is the mass of fermion $\psi$. In this article, for illustrative purposes we choose the DFSZ model to define a QCD axion band where $C_{ae} \in (0.024,1/3)$ and $|C_{aN}| \in (0.16,0.6)$. For the latter we pick, for simplicity, from the minimum and maximum absolute values of the proton and neutron couplings; see e.g.~Table I of Ref.~\cite{Irastorza:2018dyq}. We note, however, that for hadronic models like KSVZ there are no tree-level couplings to electrons (meaning $C_{ae}\sim 2\times 10^{-4}$), and in both the KSVZ and DFSZ models the uncertainties allow for $C_{aN} = 0$. So, in principle, the band could extend well below the lower limit that we will show.

The second sum in Eq.~\eqref{eq:lagrangian} describes $CP$-violating scalar interactions. In general, these will shift the minimum of the axion's potential away from its usual strong-$CP$-solving value of $\theta_{\rm eff} = 0$. These interactions can be generated by any $CP$-odd operators, as well as via higher-order $CP$-\emph{even} interactions once a small amount of $CP$-violation is introduced. Assuming there exists some small remnant angle $\theta_{\rm eff}$, the corresponding $CP$-violating axion-nucleon coupling would be~\cite{Moody:1984ba,Bertolini:2020hjc},
\begin{equation}\label{eq:scalarCPVcoupling}
g_s^N\simeq\frac{\theta_{\text {eff }}}{2f_{a}} \frac{m_{u} m_{d}}{m_{u}+m_{d}} \langle N|\bar{u} u+\bar{d} d| N\rangle \approx \theta_{\rm eff} \left(\frac{8.32 \, {\rm MeV}}{f_a} \right) \, .
\end{equation}
We have taken the nucleon [$N = (n,p)$] lattice matrix element to be~\cite{Durr:2015dna}
\begin{equation}
\frac{1}{2}\left(m_{u}+m_{d}\right)\langle N|(\bar{u} u+\bar{d} d)| N\rangle \approx 38 \, \mathrm{MeV} \, .
\end{equation}
Alternative and extended calculations have been carried out. For example, in Ref.~\cite{Bigazzi:2019hav} higher-order corrections were taken into account. Reference~\cite{Bertolini:2020hjc} performed a chiral perturbation theory calculation of the full $g_s^N$ formula, which accounts for the additional $CP$-violating contributions from meson tadpoles.

Theoretical uncertainties aside, in this article we simply wish to fix a range of values for $\theta_{\rm eff}$, to show where we expect the QCD axion to live in its $CP$-violating parameter space. Bounding this range from above is straightforward: the most recent experimental constraint~\cite{Abel:2020gbr} on the electric dipole moment of the neutron puts a tight bound of~$\theta_{\rm eff}<1.2\times 10^{-10}$ (90\% C.L.). We have used a recent lattice QCD calculation~\cite{Dragos:2019oxn} of the $\theta_{\rm eff}$ parameter from the neutron electric dipole moment (as opposed to the usual QCD sum rules~\cite{Pospelov:1999mv}), which results in a bound that accounts for both theoretical and experimental uncertainties. From below, the situation is not so clear. We would like to know the typical size of $CP$ violation to expect from the SM alone; any additional $CP$ violation coming from physics beyond the SM could then live in between the upper and lower limits of the band. SM $CP$ violation would presumably originate in the weak sector via the CKM matrix~\cite{Georgi:1986kr, Moody:1984ba}, though the precise amount is not known or easy to calculate. Previous presentations have defined an expected window for $CP$-violating couplings for QCD axion by choosing a lower bound from values between $\theta_{\rm eff}=10^{-16}$ or $10^{-14}$ (see e.g.~Refs.~\cite{Arvanitaki:2014dfa, Crescini:2016lwj}), however, these are likely to be overestimates. If we expect a small $\theta_{\rm eff}$ to originate in the weak sector we would look towards the Jarlskog invariant of the CKM matrix $V_{ij}$~\cite{Georgi:1986kr},
\begin{equation}
    J_{\rm CKM}=\operatorname{Im} V_{u d} V_{c d}^{*} V_{c s} V_{u s}^{*} \approx 3\times10^{-5} \, .
\end{equation}
Using simple dimensional analysis, it can be argued that a typical $\theta_{\rm eff}$ could be,
\begin{equation}
    \theta_{\rm eff} \sim J_{\rm CKM} G^2_F f^4_\pi \sim 10^{-18} \, .
\end{equation}
This was the argument put forward in Ref.~\cite{Georgi:1986kr}, but the result was somehow bumped up by several orders of magnitude in Ref.~\cite{Moody:1984ba} (and then adopted by experimental collaborations).
We take $\theta_{\rm eff} = 10^{-18}$ to be conservative. In terms of our $CP$-violating couplings, this results in a band of values:
 \begin{equation}
10^{-29} \left( \frac{10^9~{\rm GeV}}{f_a}\right)\lesssim g^N_s \lesssim  10^{-21}  \left( \frac{10^9~{\rm GeV}}{f_a}\right) \, .
\end{equation}

\section{Astrophysical limits} \label{sec:astrolimits}
\subsection{Electron coupling}\label{sec:electroncoupling}
The pseudoscalar axion-electron coupling $g_p^e$ allows for increased stellar energy losses by the Compton process $\gamma+e\to e+a$ and bremsstrahlung $e+Ze\to Ze+e+a$~\cite{Raffelt:1999tx,Raffelt:2006cw}.\footnote{Other process like free-bound and bound-bound transitions are less important for the cases we consider, but are important in the Sun~\cite{Redondo:2013wwa}.} These processes will accelerate the cooling of stars like red giants and white dwarfs. The excessive energy loss in red giants, for instance, will delay helium ignition, causing the mass of the stars to get larger and subsequently the tip of the red giant branch of their color-magnitude diagram to get brighter. A measurement of the brightest red giant in a globular cluster can therefore be interpreted as a bound on axionic couplings.

A recent constraint on $g_p^e$ exploiting improved distance measurements to $\omega$Cen from \emph{Gaia} finds~\cite{Capozzi:2020cbu}
\begin{equation}\label{eq:gpe-astro}
    g_p^e<1.6\times10^{-13} \quad (95\%\,{\rm C.L.}).
\end{equation}
This limit holds consistently for masses up to $m_a~\alt~10$~keV, above which emission is suppressed by threshold effects.

The red giant bounds on scalar couplings to electrons date back to the old work of Grifols and Mass\'o~\cite{Grifols:1986fc}, but were improved more recently after it was realized~\cite{Hardy:2016kme} that the resonant conversion of plasmons could lead to an additional source of cooling. The new constraint is
\begin{equation}
    g_s^e\alt 7.1 \times 10^{-16} \, .
\end{equation}
This coupling is not as relevant for the QCD axion which only interacts via a derivative coupling to the electron. Any $CP$-violation induced by a small shift in the axion's VEV will not generate a $g_p^e$, unlike the case of nucleon couplings which do couple to the axion's VEV. Hence, we only explore limits on the scalar coupling to the nucleon (see below) and not the electron. For examples of constraints on a scalar electron coupling, and its combination with pseudoscalar couplings, see e.g.~Refs.~\cite{Youdin:1996dk, Hammond:2007jm, Stadnik:2017hpa, Dzuba:2018anu, Delaunay:2017dku, Yan:2019dar}. 

Further constraints could be anticipated in the future with underground experiments looking for light scalar or pseudoscalar particles produced by the Sun~\cite{Budnik:2019olh, Bloch:2020uzh, Budnik:2020nwz}.\footnote{Though we note that any searches for light particles coupled to electrons may be complicated by the same effect that enhances their production in the Sun in the first place~\cite{Gelmini:2020xir}.}

\subsection{Nucleon coupling}
The pseudoscalar nucleon coupling, defined analogously to the electron coupling, allows for the bremsstrahlung process $N+N\to N+N+a$ in a collapsed stellar core after a supernova (SN). The neutrino events measured from SN1987A lasted for around 10~s, and thus any new mechanisms of energy loss that would accelerate this event to a shorter duration are excluded~\cite{Raffelt:1987yt}. The emission rate suffers from significant uncertainties related to post-SN accretion, core-collapse mechanisms~\cite{Bar:2019ifz}, and dense nuclear matter effects~\cite{Janka:1995ir}; not all of which were considered in detail. The SN1987A neutrino bound used in 2012 to derive the same constraints we are interested in was essentially an educated dimensional analysis~\cite{Raffelt:2006cw}. A recent revision of the bound to account for additional processes affecting the axion emissivity was presented in Ref.~\cite{Carenza_2020} (see also Refs.~\cite{Carenza:2019pxu, Carenza:2020cis}). However, there are still many uncertainties surrounding our knowledge of SN1987A which cast some doubts on how robust these neutrino bounds could be~\cite{Bar:2019ifz}.

Fortunately, we can put aside the troublesome uncertainties related to supernova neutrinos, because a comparable, but slightly more stringent bound on pseudoscalar axion-nucleon interactions was presented recently. Reference~\cite{Beznogov:2018fda} used observations of the cooling of the hot neutron star HESS J1731-347 to set,
\begin{equation}\label{eq:gpn-astro}
g_p^N<2.8\times10^{-10}\, \mbox{(90\% C.L.)}\,.
\end{equation}

The scalar nucleon interaction, on the other hand, was constrained using energy-loss arguments with globular-cluster stars through the process $\gamma+{}^4{\rm He}\to {}^4{\rm He}+a$~\cite{Grifols:1986fc, Raffelt:1999tx, Raffelt:1988gv}. The updated bound from Ref.~\cite{Hardy:2016kme} including resonant plasmon conversion is,
\begin{equation}\label{eq:gsn-astro}
g_s^N\alt 1.1\times10^{-12}\,.
\end{equation}

\subsection{Black hole spins}
The spins of astrophysical black holes can be used to rule out the existence of bosonic fields in a manner that is mostly independent of how strongly they couple to the Standard Model~\cite{Dolan:2007mj, Arvanitaki:2010sy, Pani:2012vp, Brito:2015oca, Arvanitaki:2014wva, Arvanitaki:2016qwi, Herdeiro:2016tmi}. The constraints are related to the concept of superradiance, a general term for an effect that occurs in systems with a dissipative surface possessing some angular momentum. It refers to a phenomenon in which bodies incident on a spinning surface can interact in some way and leave the system extracting some of the energy or angular momentum. In the context of black holes, one can imagine a small body entering the ergosphere of Kerr spacetime and subsequently splitting apart, thereby allowing one of the pieces to leave the system with some of the black hole's energy. This idea is also known as the Penrose process~\cite{Penrose:1971uk}. The classic ``black hole bomb" thought experiment~\cite{Press:1972zz} applies this idea to bosonic fields and takes it to the extreme: it imagines a black hole surrounded by a mirror which acts to reflect the field back after initially scattering off the black hole. The process repeats again and again, amplifying the field and eventually extracting all of the black hole's energy. 

If new light bosonic fields exist, the black hole bomb scenario is brought to reality. Perturbations in the bosonic field are excited by the Kerr spacetime~\cite{Dolan:2007mj}, and if the Compton wavelength of the field roughly matches the size of the black hole, then the boson's mass will create a confining potential, effectively acting as the mirror of the black hole bomb. If such a field exists, then excited perturbations will accumulate around the black hole and quickly act to spin it down. Therefore, the observation of any black hole spin will exclude the existence of bosonic fields over a mass range set by the black hole mass.

We use the most recent set of exclusion bounds on the masses of light bosonic fields using the set of all measured astrophysical black hole spins~\cite{Stott:2020gjj} (note that we take the 95\% C.L.~exclusion bounds found in the main text, not the 68\% C.L. reported in the abstract). The most relevant window that we consider here is the constraint from stellar-mass black holes which rule out axion masses in the window $10^{-11}$ to $10^{-14}$~eV. Though often touted as a definitive exclusion of light bosonic fields over these mass windows, these limits are somewhat model dependent. For instance, scenarios can be constructed to populate these excluded regions with light bosonic fields~\cite{Mathur:2020aqv}. There are also uncertainties related to the measurements of black hole spins which are not conservatively treated in the derivation of these bounds. Therefore this mass range grayed out in our later figures should not be treated as a definitive exclusion, but only as regions that will require more effort to understand should an axion be detected in one.

\section{Scalar nucleon interactions}\label{sec:baryon}

\begin{figure*}
\includegraphics[width=0.9\textwidth]{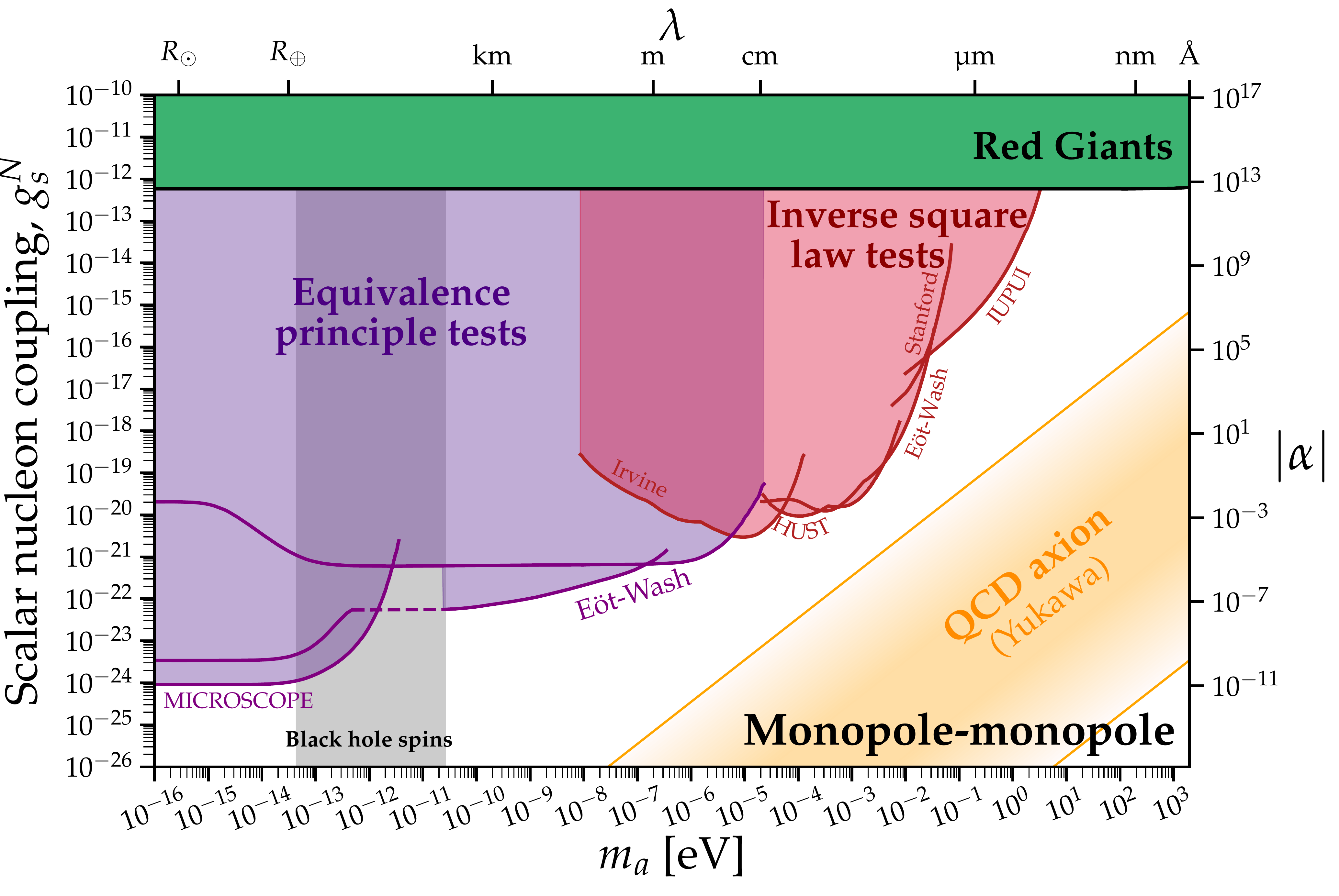}
\caption{Combined limits on a scalar nucleon coupling $g_s^N$. We express constraints in terms of the mass of the would-be axion that could mediate this $CP$-violating interaction. The range $\lambda$ and strength $\alpha$ of the forces constrained by the experimental bounds [see Eq.~(\ref{eq:NewtonYukawa})] are also indicated by the top and right-hand axes respectively. The bounds shown in red are tests of the inverse-square law~\cite{Chen:2014oda,Lee:2020zjt,Kapner:2006si,Tan:2020vpf,Hoskins:1985tn}, whereas those in purple are tests of the WEP~\cite{Smith:1999cr, Smith:1999cr, Berge:2017ovy}. In green, we show the astrophysical bound from the cooling of red giant stars~\cite{Hardy:2016kme}. In gray we show the masses disfavored by the observed spins of stellar-mass black holes~\cite{Stott:2020gjj}. The diagonal band of couplings shows roughly the expected range for the QCD axion's scalar Yukawa couplings, which we motivate in Sec.~\ref{sec:axions}. The combined bound can be downloaded from \href{https://raw.githubusercontent.com/cajohare/AxionLimits/master/limit_data/ScalarNucleon/Union.txt}{this https url}.\label{fig:scalar}}
\end{figure*}
We now consider a generic long-range monopole-monopole force mediated by any scalar (not necessarily the axion) with equal couplings to protons and
neutrons,~$g_s^N$. The Yukawa potential of Eq.~\eqref{eq:monomono} can be written as an additional term in the standard formula for Newton's gravitational potential,
\begin{equation}\label{eq:NewtonYukawa}
V=-\frac{G_{\rm N} m_1 m_2}{r}\,\left(1+\alpha\,e^{-r/\lambda}\right)\,.
\end{equation}
We can write the parameter $\alpha$ in terms of our dimensionless coupling by expressing it in terms of the atomic mass unit $m_u$, 
\begin{equation}
\alpha=\frac{\left(g_s^N\right)^2}{4\pi\,G_{\rm N} m_u^2}=
1.37\times10^{37}\left(g_s^N\right)^2\,.
\end{equation}
The range of the force is again just the inverse of the mass of the mediating particle,
\begin{equation}
\lambda=m_a^{-1}=19.73~{\rm cm}~\frac{\upmu{\rm eV}}{m_a}\,.
\end{equation}

The literature on experimental tests for these kinds of scalar-mediated forces usually show constraints in the $(\alpha,\lambda)$ parameter space. At distances above $\lambda\sim 0.1\,\upmu$m laboratory tests of Newton's inverse-square law out-compete the red giant bound. These tests dominate until around the meter scale, where the WEP probes become more viable and set the best limits down to arbitrarily light masses.

In Fig.~\ref{fig:scalar} we compile the best experimental constraints on this parameter space. We display the constraints as a function of both the parameters entering Eq.~\eqref{eq:NewtonYukawa}, ($\alpha$,$\lambda$) as well as the corresponding axion mass and scalar nucleon coupling $(m_a,g_s^N)$. The constraints shown in Fig.~\ref{fig:scalar} are described in order of increasing mass below.
\\~\\
\noindent {\bf Figure}~\ref{fig:scalar}:~\begin{itemize}
\setlength\itemsep{0em}
    \item {\bf MICROSCOPE}: a satellite-borne WEP test in orbit around the Earth, monitoring the accelerations of platinum and titanium test masses in free fall~\cite{Berge:2017ovy}.
    \item {\bf E\"ot-Wash} (purple): a group based at the University of Washington devoted to performing a range of tests of gravitational physics in the lab. The long-range sensitivity to $g_s^N$ was obtained in a WEP experiment reported in Smith et~al.~(2000)~\cite{Smith:1999cr} which measured the differential accelerations of copper and lead test bodies in a torsion balance as a 3 ton uranium attractor was rotated around them.
    \item {\bf Irvine}: tests of the inverse-square law at centimeter to meter-scales reported in Hoskins et~al.~(1985)~\cite{Hoskins:1985tn}, in which a torsion balance was used to measure torques between copper masses.
    \item {\bf HUST}: inverse-square law tests using torsion pendula at the Huazhong University of Science and Technology. The limit shown combines several reports from 2007 to 2020~\cite{Tu:2007zz,Yang:2012zzb,Tan:2020vpf,Tan:2016vwu}. The most recent of these experiments improved upon the previous limit in the sub-mm range thanks to a novel method of reducing vibrational noise on the electrostatic shielding between the test masses and the attractor.
    \item {\bf E\"ot-Wash} (red): torsion balance tests of the inverse-square law at the sub-mm to 10 micron range, presented in Kapner et al. (2007)~\cite{Kapner:2006si} and Lee et al.~(2020)~\cite{Lee:2020zjt}. The latter result mostly improved upon the 2007 bound, apart from in a very narrow window at 0.5~mm.
    \item {\bf Stanford} experiment of Geraci et al.~(2008)~\cite{Geraci:2008hb}, testing the inverse-square law at 10 micron scales with cryogenic microcantilevers. 
    \item {\bf IUPUI} Chen et al.~(2014)~\cite{Chen:2014oda}. The most competitive test of the inverse-square law at the 30--8000~nm scale comes from a differential force measurement using a microelectromechanical torsional oscillator at the Indiana University–Purdue University Indianapolis.
\end{itemize}

Many of the most competitive limits on the scalar nucleon coupling still originate from experiments using torsion balance or torsion pendulum techniques. The most notable advancements in this parameter space that we have included here are at the longest and shortest scales shown in Fig.~\ref{fig:scalar}. At the largest scales, MICROSCOPE has improved upon the previous E\"ot-Wash limits by a factor of 4 for masses below a peV. Future space-based experiments have the opportunity to extend these bounds even further in the coming years~\cite{Luo:2020rdp, Parnell:2020wwb, Lee:2019ixj}.

Tests at the sub-micron level are difficult due to the increasing prominence of vacuum fluctuations. These hinder further improvements in sensitivity, even if electrostatic backgrounds can be subtracted. The IUPUI exclusion limit shown in Fig.~\ref{fig:scalar} has advanced by over an order of magnitude from the previous limit from the same group reported in 2007~\cite{Decca:2007jq}. This is mostly thanks to a novel technique of suppressing the background from vacuum fluctuations. The technique involved coating their source mass with a film of gold thicker than the material's plasma wavelength, which acts to suppress the Casimir force between the interior of the source mass and the attractor. Tests at even smaller distances than this still currently lack the sensitivity to improve upon the astrophysical bounds~\cite{Ding:2020mic}, hence we do not show them. In the future, tests using shifts in nuclear emission lines measured with M\"ossbauer spectroscopy~\cite{Gratta:2020hyp} could potentially improve upon the sub-micron bounds.

\section{Monopole-dipole forces}\label{sec:spinbulk}
\subsection{Electron-nucleon interactions}
\begin{figure*}
\includegraphics[width=0.83\textwidth]{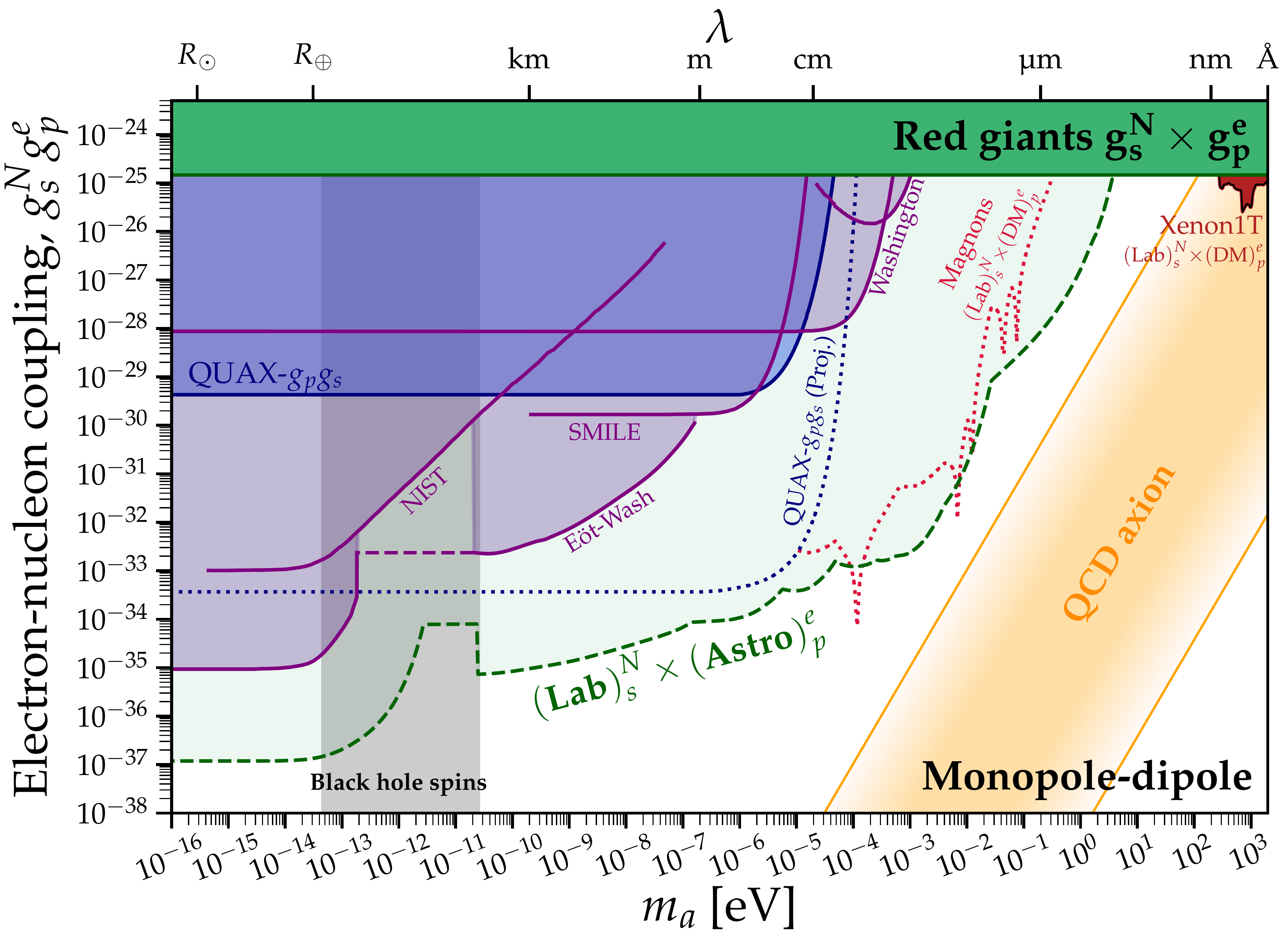}
\caption{Combined limits on the scalar-pseudoscalar nucleon-electron coupling $g_s^N g_p^e$. The filled regions bounded by solid lines are all existing laboratory bounds on this parameter space, whereas the dashed line corresponds to a combination of laboratory (scalar) and astrophysical (pseudoscalar) bounds. The dotted lines are all projections. The limit and projection for the axion search QUAX-$g_p g_s$~\cite{Crescini:2017uxs, Crescini:2016lwj} are shown in blue, whereas the limits for other searches for monopole-dipole forces between nucleons and electrons are shown in purple~\cite{Wineland:1991zz, Heckel:2008hw, Lee:2018vaq, Terrano:2015sna, Hoedl:2011zz}. We also show the limit from the Xenon1T dark matter search~\cite{Aprile:2019xxb} (multiplied by the astrophysical bound on $g_s^N$ from the previous figure). We also show a projection for a proposed axion dark matter detector based on magnons that is sensitive to $g_p^e$ (also multiplied by the experimental bound on $g_s^N$). The combined astrophysical and laboratory bound can be downloaded from \href{https://raw.githubusercontent.com/cajohare/AxionLimits/master/limit_data/MonopoleDipole/ElectronNucleon/UnionAstroLab.txt}{this https url}.
\label{fig:ne}}
\end{figure*}
We now come to constraints on the monopole-dipole interaction for the combination of the scalar nucleon coupling and the pseudoscalar electron coupling. A summary of these bounds is shown in Fig.~\ref{fig:ne}. The most restrictive limit on $g_s^N g_p^e$ arises from the long-range force limits on $g_s^N$ shown in Fig.~\ref{fig:scalar} and the astrophysical $g_p^e$ limit from Eq.~\eqref{eq:gpe-astro}. This combined bound is shown as a green dashed line: currently none of the existing or projected constraints are sufficient to improve upon it substantially. The purely astrophysical red giant bound in Fig.~\ref{fig:ne} is found by multiplying Eqs.~\eqref{eq:gpe-astro} and~\eqref{eq:gsn-astro}.

As well as laboratory searches for monopole-dipole forces (shown in purple), we also show (in blue) the limit and projection for QUAX-$g_p g_s$: an experiment devoted, at least nominally, to probing axions~\cite{Crescini:2017uxs}. Several experiments fall under the umbrella of QUAX; the limits shown here are for a setup that is similar in design to the proposed ARIADNE (which we discuss in the next section). The concept aims to search for an axion-mediated force in between unpolarized nucleons and polarized electrons. The unpolarized nucleons take the form of small lead masses which are placed at regular intervals on the edge of a spinning wheel. This wheel is spun at a distance of a few centimeters from a small crystal of paramagnetic gadolinium orthosilicate (GSO). The axion field sourced by the lead masses would induce a varying magnetization signal in the crystal with a frequency given by the rate at which the masses pass by the polarized sample. With an RLC circuit tuned to this frequency, the oscillating magnetization signal could then also be amplified. We take the current exclusion limit from QUAX's first $g_p g_s$ experiment from Ref.~\cite{Crescini:2017uxs}, and their resonant RLC projection from Ref.~\cite{Crescini:2016lwj}.

The constraints on $g_s^N g_p^e$ are described in more detail below, ordered from low to high masses.
\\~\\
\noindent {\bf Figure}~\ref{fig:ne}:
\begin{itemize}
\setlength\itemsep{0em}
    \item {\bf E\"ot-Wash} experiment reported in Heckel et~al.~(2008)~\cite{Heckel:2008hw} with a spin pendulum made of two materials containing a high density of polarized electrons, and the Earth and Sun as source masses.
    \item {\bf NIST}: A stored-ion spectroscopy experiment on $^9$Be$^+$ atoms by Wineland et~al.~(1991)~\cite{Wineland:1991zz} in which the Earth played the role of the source mass.
    \item {\bf SMILE}: probing forces between polarized electrons in a $^3$He-K comagnetometer, and unpolarized lead weights spaced 15 cm away~\cite{Lee:2018vaq}.
   \item {\bf QUAX}-$g_p g_s$ exclusion limit with a 1 cm$^3$ sample of GSO~\cite{Crescini:2017uxs}.\footnote{We note that there seems to be an issue with the QCD band shown in the exclusion plots of Refs.~\cite{Crescini:2016lwj,Crescini:2017uxs} which is several orders of magnitude too high in coupling, and only scales with $m_a$ instead of $m^2_a$.}
       \item {\bf QUAX}-$g_p g_s$ projection for their sensitivity amplified with a resonant RLC circuit~\cite{Crescini:2016lwj}.
    \item {\bf Washington} limits from two experiments using polarized torsion pendula: Terrano et~al.~(2015)~\cite{Terrano:2015sna} and Hoedl et~al.~(2011)~\cite{Hoedl:2011zz}.
\item {\bf Magnon}-based axion dark matter search for the axion-electron coupling. We show the projection from Ref.~\cite{Mitridate2020}, though another related proposal was made in Ref.~\cite{Chigusa:2020gfs}. This idea is conceptually very similar to the QUAX Collaboration's proposed dark matter search~\cite{Crescini:2018qrz, Ruoso:2015ytk}. 
    \item {\bf Xenon1T}'s underground dark matter axion search for keV electron recoils~\cite{Aprile:2019xxb}.
\end{itemize}
Note that the XENON1T and Magnon projections are for dark matter experiments and involve a multiplication by the monopole-monopole constraint $g_s^N$ from Fig.~\ref{fig:scalar}. Even accounting for projections, no proposed experiment is yet sufficient to break through into the corner of parameter space in which the QCD axion could live.

\subsection{Nucleon-nucleon interactions}
\begin{figure*}
\includegraphics[width=0.83\textwidth]{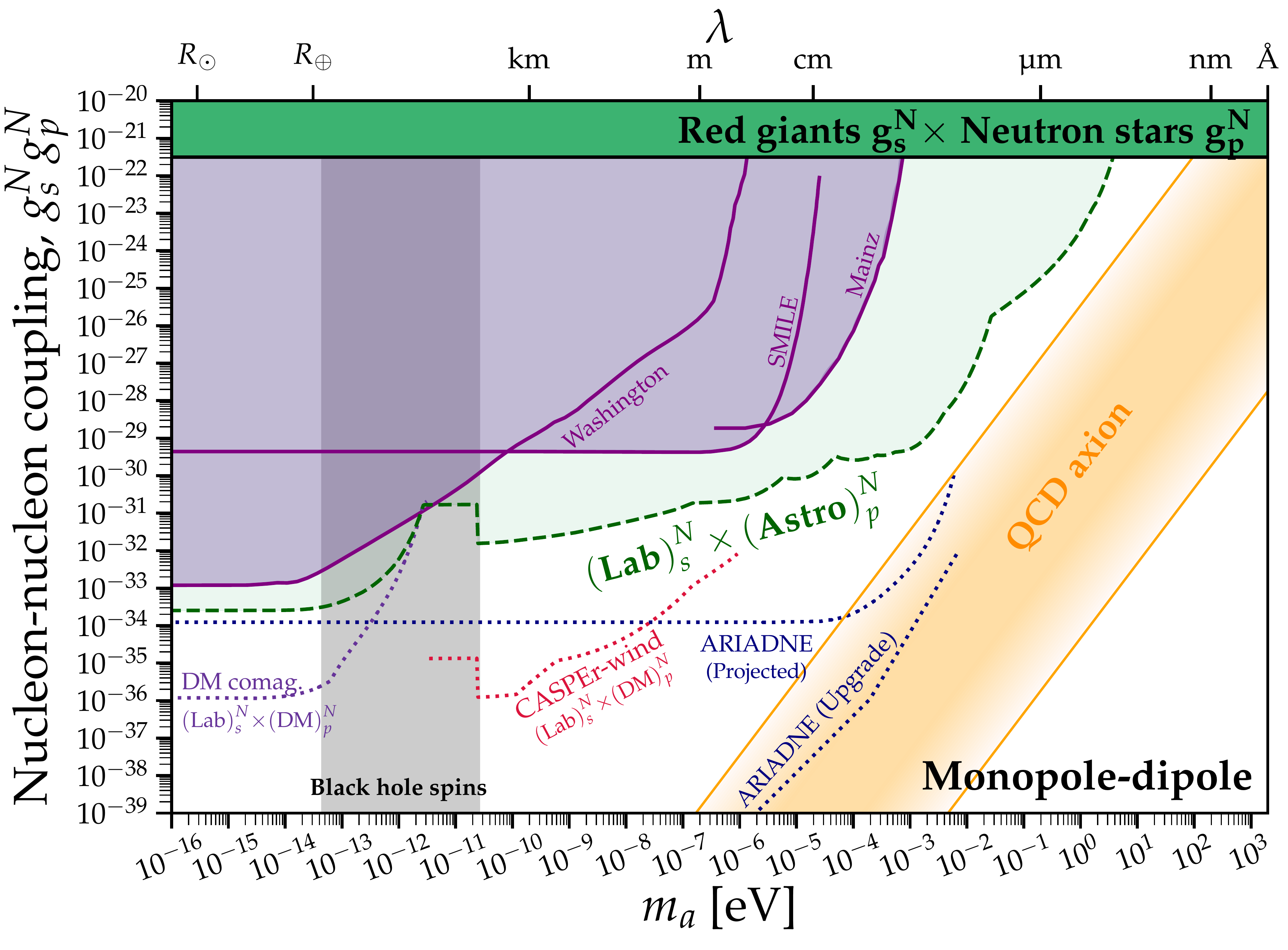}
\caption{Upper limits on $g_s^N g_p^N$. The solid lines are all existing limits on this parameter space, the dashed lines correspond to a combination of laboratory scalar searches and astrophysical pseudoscalar bounds, and the dotted lines are all projections. The two projections for ARIADNE~\cite{Arvanitaki:2014dfa} aim to have QCD sensitivity for $10 \upmu$eV--meV axion masses. We also show projected limits for dark matter experiments: CASPEr-wind~\cite{JacksonKimball:2017elr}, and a possible future dark matter comagnetometer~\cite{Bloch:2019lcy}. In both of these cases we have multiplied the expected constraint on $g_p^N$ with the astrophysical bound on $g_s^N$. The combined astrophysical and laboratory bound can be downloaded from \href{https://raw.githubusercontent.com/cajohare/AxionLimits/master/limit_data/MonopoleDipole/NucleonNucleon/UnionAstroLab.txt}{this https url}.\label{fig:nn}}
\end{figure*}
Similar to the electron-nucleon interaction, the most stringent limit on $g_s^N g_p^N$ can be derived by multiplying the long-range force limits shown Fig.~\ref{fig:scalar} with the neutron star cooling bound on the pseudoscalar coupling written in Eq.~\eqref{eq:gpn-astro}. We show these bounds in Fig.~\ref{fig:nn}. As in the previous example, we show the combination of the lab bound on the scalar coupling with the astrophysical bound on the pseudoscalar coupling with a green dashed line. The three most stringent purely experimental bounds are described below.
\\~\\
\noindent {\bf Figure}~\ref{fig:nn}:~\begin{itemize}
\setlength\itemsep{0em}
    \item {\bf Washington} experiment of Venema et al.~(1992)~\cite{Venema:1992zz} which measures the spin precession frequencies of two Hg isotopes optically, using the Earth as a source mass. Note that we have taken the version of this limit presented in Fig.~13 of Ref.~\cite{Safronova:2017xyt}.
    \item {\bf SMILE} experiment probing forces between polarized nucleons in a $^3$He-K comagnetometer, and unpolarized lead weights spaced 15 cm away~\cite{Lee:2018vaq}.
    \item {\bf Mainz} experiment~\cite{Tullney:2013wqa} using an ultra-sensitive low-field magnetometer with polarized gaseous samples of $^{3}$He and $^{129}$Xe.
\end{itemize}
We also highlight two potential dark matter limits coming from experiments sensitive to $(g_p^N)^2$: the upcoming nuclear magnetic resonance experiment CASPEr-wind~\cite{JacksonKimball:2017elr}, and a concept for a dark matter comagnetometer suggested by Ref.~\cite{Bloch:2019lcy}.

One of the most notable updates since the last compilation of these bounds was presented is the first limit mentioned above~\cite{Venema:1992zz}. Reference~\cite{Raffelt:2012sp} did not consider bounds at scales larger than 10 m for this interaction, so extending our scope to larger scales means that the constraints at the lightest masses have improved by around five orders of magnitude. Some experimental techniques probing around 0.01~eV have also improved since the last compilation, e.g.~from experiments using ultracold neutrons~\cite{Afach:2014bir}, and hyperpolarized $^3$He~\cite{Guigue:2015fyt}. However these limits do not yet reach the purely astrophysical bounds hence we do not show them.

The most interesting projection in this space (and for all the parameter spaces we show here) is the proposed experiment ARIADNE. This proposed experiment based at the University of Nevada, Reno~aims for sensitivity well into the QCD band~\cite{Arvanitaki:2014dfa, Geraci:2017bmq}. If successful in meeting its projections, ARIADNE will be the only purely laboratory search with sensitivity better than any lab$\times$astro combination. The general concept is similar to QUAX-$g_p g_s$ discussed earlier. ARIADNE will consist of a spinning unpolarized source mass with teeth that extend radially outwards towards a fixed laser-polarized $^3$He detector. The source mass is spun so that the teeth pass by the detector at the spin-precession frequency. The resonantly enhanced transverse magnetization induced by an axion-mediated monopole-dipole force can then be read out with a SQUID, assuming magnetic backgrounds can be shielded sufficiently~\cite{Fosbinder-Elkins:2017osp}. Both curves shown in Fig.~\ref{fig:nn} assume a $10^6$~s integration time. However, the sensitivity will be limited by the relaxation time of the $^3$He sample. The upper curve is the projection for ARIADNE's first stage~\cite{Arvanitaki:2014dfa}, assuming a relaxation time of 1000~s. The lower curve is what could be anticipated in the future for a scaled-up version. 

\section{Dipole-dipole forces} \label{sec:dipoledipole}

Dipole-dipole forces dependent on $(g_p^e)^2$ and $(g_p^N)^2$ can also be searched for in the laboratory. A recent summary of experimental bounds can be found in, for example, Ref.~\cite{Mostepanenko:2020lqe}. Unfortunately, the results are much less restrictive than the corresponding astrophysical limits by many orders of magnitude. 

For the nucleon coupling the astrophysical bound is at the $g_p^N \sim 10^{-10}$ level [see Eq.~\eqref{eq:gpn-astro}], whereas even one of the most restrictive experimental limits, the Princeton K--$^3$He comagnetometer~\cite{Vasilakis:2008yn}, only sets a bound of $g_p^N<4.6\times10^{-5}$ below $m_a\lesssim$~meV. At higher masses the constraints are even weaker~\cite{Adelberger:2006dh}. At much lower masses the CASPEr ultralow magnetic field spin precession experiments~\cite{Garcon:2019inh, Wu:2019exd}, and a possible proton storage ring experiment~\cite{Graham:2020kai} are at a similar level of sensitivity in coupling.

For the electron pseudoscalar coupling $g_p^e$, the red giant and white dwarf bounds are competitive across all relevant masses, up to heavy keV-scale axions which can be probed more sensitively by underground dark matter searches~\cite{Aprile:2019xxb}. Future underground detectors like the multiton xenon time projection chamber DARWIN will extend the reach for these high masses~\cite{Aalbers:2016jon}, and various semiconductor and solid-state detectors could extend the reach for sub-keV dark matter axions~\cite{Bloch:2016sjj, Griffin:2020lgd, Armengaud:2018cuy}. Again, these constraints all rely on heavy axions comprising a decent fraction of the dark matter.

In principle, underground detectors are also sensitive to $g_p^e$ down to arbitrarily low masses because they can detect the flux of solar axions, also at keV energies. However given the fact that the event rate scales with $(g_p^e)^4$, this will require experiments with kton-year exposures to even reach values like $g_p^e \sim 10^{-13}$. A solar axion search has been conducted for the pseudoscalar axion-nucleon coupling as well~\cite{Bhusal:2020bvx}. Since the stellar bounds are so stringent in these cases, it is unlikely that any experimental probe will be able to improve upon these bounds unless it is a search reliant on axions comprising dark matter.

So far the only axion haloscope experiments that have been proposed for the axion-electron coupling are the designs sketched in Refs.~\cite{Chigusa:2020gfs, Mitridate2020} (shown in Fig.~\ref{fig:ne}), which aim to couple the axion to magnons and polaritons in condensed matter systems. However, these proposals need further analysis to prove their sensitivity.


Another possibility for the future is the various proposals for the detection of dark matter dipole-dipole couplings to nucleons and electrons with spin precession techniques. One burgeoning field mentioned in Ref.~\cite{Graham:2017ivz} that we wish to highlight is atom interferometry, as several proposals are already underway. Some examples include the meter to km-scale interferometers like AGIS~\cite{Dimopoulos:2008sv}, AION~\cite{Badurina:2019hst}, MAGIS~\cite{Graham:2017pmn}, MIGA~\cite{MIGA}, ELGAR~\cite{Canuel:2019abg}, and ZAIGA~\cite{ZAIGA}, as well as a proposed space-based experiments~\cite{Aguilera:2013uua,Dimopoulos:2008sv,Tino:2019tkb,Bertoldi:2019tck}. Coordination between several globally distanced interferometers has also been suggested~\cite{Badurina:2019hst}. 

Having already been proven their utility since the 1990s as viable gravitational gradiometers~\cite{Snadden:1998zz}, tests of the WEP~\cite{Barrett}, Lorentz invariance~\cite{Muller:2007es}, and for measuring inertial forces~\cite{Canuel:2006zz}; atom interferometers are of particular relevance currently as they can also serve as gravitational-wave detectors. The proposals of Refs.~\cite{Badurina:2019hst, Graham:2017pmn} in particular target the mid-band (30 mHz to 10 Hz) gap in frequency sensitivity between LISA and LIGO. In addition to gravitational waves, interferometry experiments have been suggested for the detection of light scalar and vector dark matter candidates~\cite{Geraci:2016fva, Arvanitaki:2016fyj}. 

An atom interferometry experiment as a dark matter detector would work by collecting the phases accumulated by two ultracold atomic clouds as they travel along two very similar spatial paths. The experiment could operate in a resonant mode, similar to Ref.~\cite{Graham:2016plp}, if the atomic clouds were addressed regularly with a laser pulse, flipping their spins with a frequency matching the axion mass. Such an experiment would be able to gain sensitivity to pseudoscalar nucleon-nucleon interactions superseding the astrophysical bounds, but only if axions lighter than $m_a\sim 10^{-15}$~eV comprised the majority of the dark matter~\cite{Graham:2017ivz}.

\section{The need for purely laboratory searches}\label{sec:astrolabbounds}
The bounds we have presented here are the most restrictive ones to date on these couplings. However, we caution that they rely on the combination of laboratory and astrophysical constraints, with each set using very different methodologies. While laboratory constraints can be regarded as essentially robust with statistically rigorous definitions, astrophysical constraints often come parceled with possibly unwanted uncertainties. For instance, the previously used bounds on axion couplings from the neutrino burst of SN1987A have been the subject of some questioning recently~\cite{Bar:2019ifz, Carenza:2019pxu}. 

On the other hand, an argument in favor of astrophysical bounds in general can be made by realizing that many similar bounds can be derived using a variety of different data sets. Here, we have simply reported the most stringent ones, namely the cooling of a particular neutron star for $g_p^N$~\cite{Beznogov:2018fda}, the red giant branch branch of $\omega$Cen for $g_p^e$~\cite{Capozzi:2020cbu}, and M5 for $g_s^N$~\cite{Viaux:2013lha,Hardy:2016kme}. However, other constraints exist, such as those using the cooling of white dwarfs~\cite{Hansen_RG,Bertolami:2014wua}, other sets of neutron stars~\cite{Sedrakian:2015krq}, and other globular clusters~\cite{Raffelt:1988gv,Ayala:2014pea,Capozzi:2020cbu} (see Refs.~\cite{Giannotti:2017hny,DiLuzio:2020jjp} for recent work that combines different bounds). Put together, the existence of astrophysical bounds across axion masses below the keV scale are robust to at least the order of magnitude, if not at additional significant figures.

What does complicate matters, however, is if any new physics takes place in astrophysical environments in a way that could spoil the astrophysical bounds.  There is a history of such scenarios being proposed, usually inspired by surprising experimental hints that were ostensibly in conflict with more stringent astrophysical bounds. Most notable in this regard are the PVLAS observation of photon polarization rotation from 2005~\cite{Zavattini:2005tm}, and XENON1T's more recent observed excess of electronic recoils with a spectrum resembling that of solar axions~\cite{XENON1T}.\footnote{The reason behind the former observation was ultimately determined to be a newly discovered experimental systematic~\cite{Zavattini:2007ee}; the true origin of the XENON1T excess remains to be seen.} These generally involve introducing a mechanism by which the additional cooling of stars by axionic emission is suppressed~\cite{Jain:2005nh, Masso:2005ym, Jaeckel:2006xm, Masso:2006gc, Budnik:2020nwz, DeRocco:2020xdt, Bloch:2020uzh}. One challenge in developing these scenarios is to explain the apparent ``chameleonic'' environment dependence, i.e. why emission is different in red giants, white dwarfs, or our Sun, even when the emission mechanisms and energy scales are comparable.

To give one recent example, Ref.~\cite{DeRocco:2020xdt} constructed a simple model that includes a new scalar field and two vector-like fermions coupled to the axion. The key feature of the model is that the scalar field has a VEV sourced by the local baryonic density. Then via the fermion couplings to this VEV, the mechanism ultimately gives rise to an axion mass which also varies with density. This latter example is similar in spirit to Ref.~\cite{Budnik:2020nwz} in that it aims to arrange the new degrees of freedom to adjust the axion mass; whereas other attempts focused on environment-dependent couplings~\cite{Bloch:2020uzh, Masso:2005ym, Jaeckel:2006xm, Masso:2006gc}.

Although the scenarios we have mentioned do not necessarily have the most solid of theoretical motivations, they can nevertheless be conjured in quite generic and straightforward ways (as long as one admits a bit of light fine-tuning). This is perhaps cause for concern if we are going to rely on astrophysical bounds to guide us towards the corners of parameter space where we want future experiments to search. Evidently, if a scenario like the ones we have mentioned is true, then any combination of laboratory and astrophysical bounds is overly stringent and could lead to a premature abandoning of the axion as an attractive theoretical target. Additionally, if the axion did change its properties with its environment like a chameleon, then understanding this complex phenomenology is going to be challenging if all we have are astrophysical probes.

We therefore need future experiments on Earth. However, searches based on the axion's role as dark matter are even more fraught. Axions need not comprise even a subdominant fraction of the dark matter density in the galaxy (relied upon by some searches~\cite{Aprile:2019xxb, Bloch:2019lcy, JacksonKimball:2017elr, Mitridate2020, Chigusa:2020gfs, Griffin:2020lgd}), and even if they do there are hefty astrophysical uncertainties on their local distribution~\cite{OHare:2017yze, Dokuchaev, OHare:2018trr, Knirck:2018knd, OHare:2019qxc, Evans:2018bqy}. The only way to truly confirm or rule out the existence of a new weakly coupled particle like the axion will be to perform purely laboratory searches.

\section{Summary}\label{sec:conclusions}
\begin{figure}
\includegraphics[width=\columnwidth]{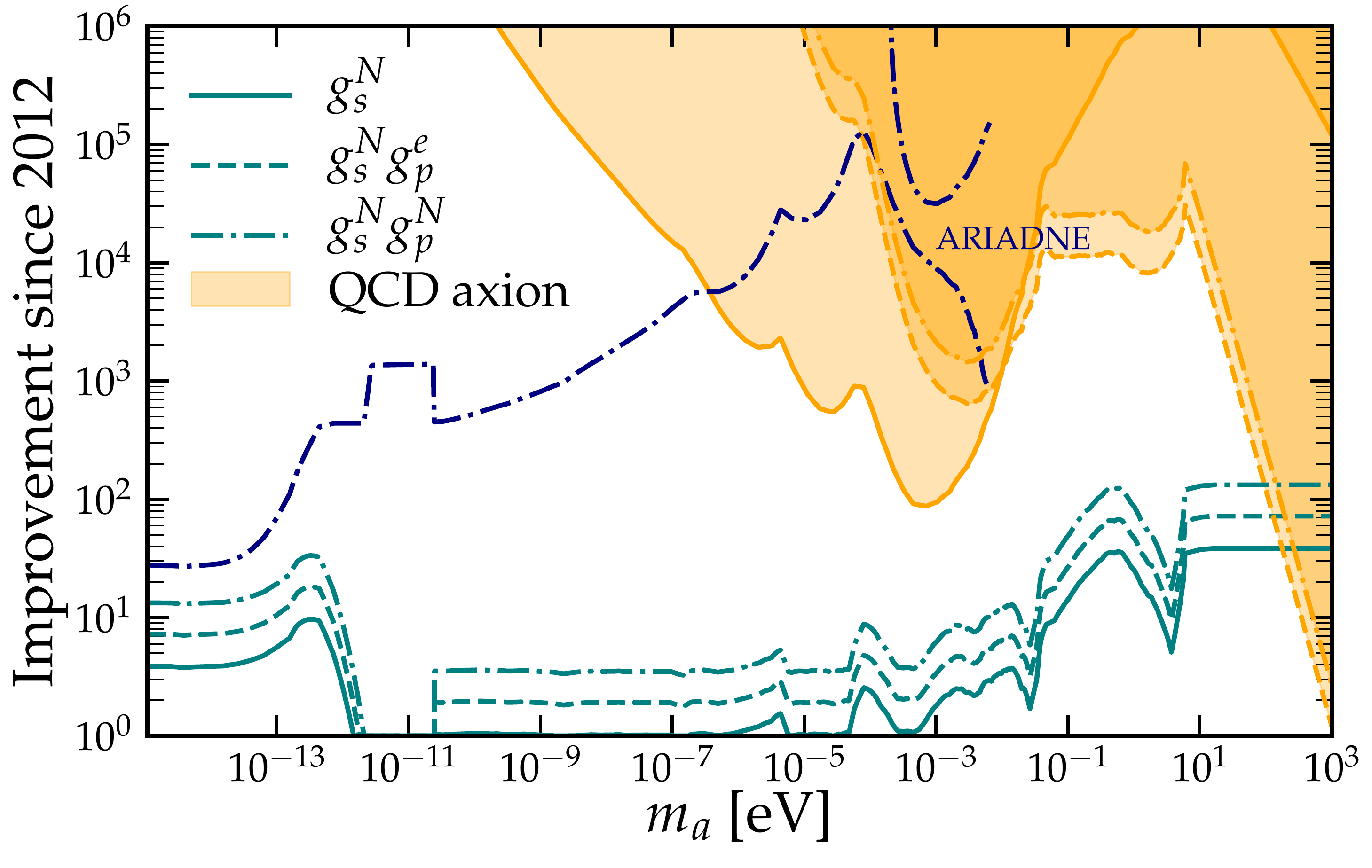}
\caption{Improvement in the constraints on the three couplings studied in this work. We calculate the improvement factor for each coupling by taking our new limit and dividing by the previously combined limit from 2012~\cite{Raffelt:2012sp}. The different line styles correspond to the three couplings: scalar nucleon monopole-monopole coupling, $g_s^N$ (solid); nucleon-electron monopole-dipole coupling $g_s^N g_p^e$ (dashed); and nucleon-nucleon monopole-dipole coupling $g_s^N g_p^N$ (dot-dashed). In orange, and with the same line style, we show the \emph{required} improvement (relative to the 2012 limit) that would be needed to reach QCD sensitivity. The blue dot-dashed lines correspond to the sensitivity improvement projected for the two ARIADNE estimates shown in Fig.~\ref{fig:nn}. For the scalar, electron-nucleon, and nucleon-nucleon couplings, we achieve maximum enhancement of factors of 40, 70 and 130 in sensitivity, respectively.\label{fig:improvement}}
\end{figure}
We have revised the experimental and astrophysical bounds on the $CP$-violating couplings expected to be present in QCD axion models. Relative to the previous compilation from 2012~\cite{Raffelt:2012sp} we saw improvements of up to a factor of 40 for the scalar coupling to nucleons, 70 for the monopole-dipole nucleon-electron coupling, and 130 for the monopole-dipole nucleon-nucleon coupling. The improvement factors as functions the axion mass are shown in Fig.~\ref{fig:improvement}. We also show the improvement still required to reach the expected levels of $CP$-violation in QCD axion models~\cite{Moody:1984ba, Georgi:1986kr}.

All of the coupling combinations studied here have benefited from the improved astrophysical limits, affecting all masses equally below $m_a \sim$~MeV. These improvements have arisen thanks to more accurate distances to globular clusters thanks to \emph{Gaia}~\cite{Capozzi:2020cbu}, a new analysis of neutron star cooling~\cite{Beznogov:2018fda}, and refined calculations of scalar-induced cooling mechanisms in red giants~\cite{Hardy:2016kme}. The most significant mass-dependent improvement is for axions above $m_a\sim$~meV: mostly thanks to the new inverse-square law test at IUPUI~\cite{Chen:2014oda} which has improved the bound on $g_s^N$ at sub-micron scales.

Of the currently published projections for future laboratory searches, the most hotly anticipated one will be ARIADNE~\cite{Arvanitaki:2014dfa} which aims for sensitivity well into the QCD band of $g_s^N g_p^N$ provided there are no close cancellations in the constituent couplings. We show these expected improvements in Fig.~\ref{fig:improvement} as well. If we include dark matter searches as well, we would also see a dramatic improvement in the constraints on keV-scale axions from underground particle detectors like XENON1T~\cite{Aprile:2019xxb}.
Future dark matter searches for the axion's dipole-dipole nucleon coupling with CASPEr-wind~\cite{JacksonKimball:2017elr} or with a possible dark matter comagnetometer~\cite{Bloch:2019lcy} could improve the limits for masses below the peV scale. 
We have stressed in Sec.~\ref{sec:astrolabbounds} that some of these limits must be regarded with caution as they combine different constraints from laboratory experiments and astrophysical bounds. Many astrophysical bounds come with uncertainties, and if there is any environment-dependent new physics linked to the axion sector, this could potentially frustrate our ability to set astrophysical bounds at all. Ultimately the only way to truly confirm or rule out the existence of a new weakly coupled particle like the axion will be to perform purely laboratory searches. In light of these extremely small $CP$-violating couplings, future experiments like ARIADNE and QUAX are perhaps even more crucial. 

\section*{Acknowledgements}
We thank Luca Di Luzio, Maxim Pospelov, Jordy De Vries for enlightening correspondence, and Georg Raffelt for highly useful comments on the manuscript. CAJO is supported by the University of Sydney and the Australian Research Council. EV acknowledges support by the US Department of Energy (DOE) Grant No.  DE-SC0009937.

\bibliographystyle{bibi.bst}
\bibliography{AxionCPV.bib}

\end{document}